\documentclass[aps,amssymb,prb,twocolumn,superscriptaddress]{revtex4}
\usepackage{graphicx}
\begin{document}
\title{Phonon thermal conductivity in doped $\rm\bf La_2CuO_4$: relevant scattering mechanisms}

\author{C. Hess}
\email[]{christian.hess@physics.unige.ch}

\affiliation{D\'{e}partment de Physique de la Mati\'{e}re Condens\'{e}e, Universit\'{e} de Gen\`{e}ve, Gen\`{e}ve,  Switzerland}
%\affiliation{2. Physikalisches Institut, RWTH-Aachen, 52056 Aachen, Germany}
%\affiliation{II. Physikalisches Institut, Universit\"{a}t zu K\"{o}ln, 50937 K\"{o}ln, Germany}

\author{B. B\"{u}chner}

\affiliation{2. Physikalisches Institut, RWTH-Aachen, 52056 Aachen, Germany}

\author{U. Ammerahl}
%\affiliation{II. Physikalisches Institut, Universit\"{a}t zu K\"{o}ln, 50937 K\"{o}ln, Germany}
\affiliation{Laboratoire de Physico-Chimie, Universit\'{e} Paris-Sud, 91405 Orsay, France}

%\affiliation{}

\author{A. Revcolevschi}
\affiliation{Laboratoire de Physico-Chimie, Universit\'{e} Paris-Sud, 91405 Orsay, France}

\date{\today}

\begin{abstract}
Results of in-plane and out-of-plane thermal conductivity measurements on $\rm La_{1.8-x}Eu_{0.2}Sr_xCuO_4$ ($0\leq x\leq0.2$) single crystals are presented. The most characteristic features of the temperature dependence are a pronounced phonon peak at low temperatures and a steplike anomaly at $T_{LT}$, i.e., at the transition to the low temperature tetragonal phase (LTT-phase), which gradually decrease with increasing Sr-content. Comparison of these findings with the thermal conductivity of $\rm La_{2-x}Sr_xCuO_4$ and $\rm La_2NiO_4$ clearly reveals that in $\rm La_{2-x}Sr_xCuO_4$ the most effective mechanism for phonon scattering is impurity-scattering (dopants), as well as scattering by soft phonons that are associated with the lattice instability in the low temperature orthorhombic phase (LTO-phase). There is no evidence that stripe correlations play a major role in suppressing the phonon peak in the thermal conductivity of $\rm La_{2-x}Sr_xCuO_4$.
\end{abstract}

\pacs{}
%\keywords{}

\maketitle

%%%%%%%%%%%%%%%%%%%%%%%%%%%%%%%%%%%%%%%INTRODUCTION%%%%%%%%%%%%%%%%%%%%%%%%%%%%%%%%%%%%%%%%%%%
\section{\label{intro}Introduction}
The segregation of spins and holes into stripelike arrangements appears to be a common feature of
doped Mott insulators.\cite{Tranquada94,Tranquada95,Tranquada95a,Lee97,Mori98} These so-called {\em
stripe correlations} seem to be of particular importance in understanding the electronic phase
diagram of high-temperature superconductors, where a competition between a static stripe phase and
the superconducting phase is widely discussed.\cite{White99,Chakravarty01,Zaanen03} Such a
competition is expected to be reflected by the dynamics of stripes, i.e., static stripes should
reduce the superconducting order parameter while stripes should be fluctuating in the presence of
fully developed superconductivity. Indeed, there is growing experimental evidence for such a
scenario.\cite{Buchner94,Tranquada95,Klauss00,Lake01,Lake02} While many experiments give evidence
towards static stripes of holes and spins,\cite{Tranquada95,Tranquada97,Tranquada99} signatures of
stripe fluctuations currently only comprise magnetic correlations.\cite{Cheong91,Yamada98,Mook98} A
direct observation of fluctuating charge stripes, however, is still missing. A promising
alternative approach to study stripe dynamics involves the phonon heat transport which is an
indirect probing method. Since charge stripes lead to lattice distortions, a sensitivity of the
phonon heat transport to the dynamics as well as the degree of periodicity of stripes can be
expected.

The thermal conductivity $\kappa$ of doped $\rm La_2CuO_4$ has repeatedly been the subject of
experimental research, yet no detailed understanding of the rather complicated and strong changes
of the temperature dependence of $\kappa$ upon partial substituting Sr and/or small Rare Earths (RE) like
Nd or Eu for La has been achieved. For example, in the antiferromagnetic insulators $\rm
La_{2-y}(RE)_yCuO_4$ a huge peak at room temperature which arises due to magnetic heat transport is
found in the thermal conductivity parallel to the $\rm CuO_2$-planes ($\kappa_{ab}$), while the
thermal conductivity perpendicular to the $\rm CuO_2$-planes ($\kappa_c$) is purely phononic
without a high temperature peak.\cite{Hess03} Similarly intriguing is a strong suppression of
the phononic low temperature peak in both $\kappa_{ab}$ and $\kappa_c$, which is found in the
superconducting doping levels of $\rm La_{2-x}Sr_xCuO_4$.\cite{Nakamura91} There is a seeming
correlation between this suppression and superconductivity because a phononic low temperature peak
reappears in overdoped, non-superconducting $\rm La_{2-x}Sr_xCuO_4$ as well as in $\rm
La_{2-x-y}(RE)_ySr_xCuO_4$, where superconductivity is suppressed in favor of static stripe
order.\cite{Buchner94,Baberski98,Klauss00} Baberski et al.~\cite{Baberski98} qualitatively
explained these observations based on the idea that in superconducting compounds fluctuating
stripes provide a new scattering channel for phonons. In recent studies by Sun et al., such a scattering channel plays an important role in the data interpretation.\cite{Sun03,Sun03a,Sun03b}

There is clear-cut evidence that in isostructural stripe ordering $\rm La_{2-x}Sr_xNiO_4$ the
phonon thermal conductivity $\kappa_{\mathrm{ph}}$ is closely correlated with both the dynamics and
the periodicity of stripes: While $\kappa_{\mathrm{ph}}$ is almost unaffected in the presence of
static and long range ordered stripes, it is strongly suppressed as soon as the stripes become
disordered or dynamic.\cite{Hess99,Cassel99a,Lee02} Apparently, in these compounds the thermal
conductivity is indeed a probe for stripe correlations. One might question, however, whether this
is true also in the cuprates for two reasons: First, the electron-phonon coupling in the nickelates
is much stronger than in the cuprates.\cite{Eisaki92,Anisimov92,Bi93,Chen93,Zaanen94,Cheong94}
Therefore, the effect of stripes on $\kappa_{\mathrm{ph}}$ can be expected to be much smaller in
the cuprates. Second, from a structural point of view the situation in the nickelates is much
simpler than in the cuprates. In the latter a structural instability exists and as a consequence a
number of structural phase transitions occur as a function of temperature as well as of Sr- and
RE-content.\cite{Buchner94} Since the structural instability involves soft phonon
modes,\cite{Birgeneau87,Pintschovius90,Martinez91,Keimer93} enhanced scattering of the heat
carrying phonons is likely.
%since $\kappa_{\mathrm{ph}}$ of non-doped $\rm La_2CuO_4$ exceeds $\kappa_{\mathrm{ph}}$ of all doped compounds and, in particular, exhibits a pronounced low temperature phonon peak.\cite{Nakamura91} Further,

%%%%%%%%%%%%%%%%%%%%%%%%%%%%%%%%%%%%%%%SHORT DESCRIPTION%%%%%%%%%%%%%%%%%%%%%%%%%%%%%%%%%%%%%%
In this article we reinvestigate the phonon thermal conductivity $\kappa_{\mathrm{ph}}$ of doped
$\rm La_{2-x}Sr_xCuO_4$ and present new experimental results on Eu-doped single crystals. The
single crystalline data allow us to investigate the anisotropy of the $\kappa$-tensor and provide
more precise absolute values of $\kappa$. In previous measurements on polycrystals\cite{Sera97,Baberski98} this information was not available. It is, however, necessary in order to judge the strength and therefore the importance of various scattering mechanisms for phonons. The analysis of our data yields
compelling arguments that both impurities (dopants) and soft phonons, which are associated with the
lattice instability in these compounds, strongly scatter phonons and therefore must not be
neglected in the data interpretation. Data on $\rm La_2NiO_4$ corroborate this conclusion and allow
us to qualitatively understand $\kappa_{\mathrm{ph}}$ of $\rm La_{2-x-y}(RE)_ySr_xCuO_4$ in a wide
doping range. In particular, it is not necessary to include a stripe induced scattering channel.
%The detailed discussion of the magnetic contributions and contributions to $\kappa$ due to optical phonons is subject of forthcoming papers.\cite{Hess03,Hess03b}

%%%%%%%%%%%%%%%%%%%%%%%%%%%%%%%%%%%%%STRUCTURE%%%%%%%%%%%%%%%%%%%%%%%%%%%%%%%%%%%%%%%%%%%%%%%%
The structure of this paper is as follows: After a brief description of the experimental details in section \ref{exp}, we review previous results on the thermal conductivity of doped $\rm La_{2}CuO_4$ in section \ref{lacuo}, before we proceed to the presentation of our new experimental results on $\rm La_{1.8-x}Eu_{0.2}Sr_xCuO_{4}$ in section \ref{data}. Our main results will be discussed in section \ref{discussion}.

%%%%%%%%%%%%%%%%%%%%%%%%%%%%%%%%%%%%EXPERIMENTAL%%%%%%%%%%%%%%%%%%%%%%%%%%%%%%%%%%%%%%%%%%%%%%%%%%%
\section{\label{exp}Experimental}
We have prepared single crystals of $\rm La_{1.8-x}Eu_{0.2}Sr_xCuO_{4}$ ($x=0$, 0.08, 0.15, 0.2) as
well as of $\rm La_2NiO_4$ utilizing the traveling solvent floating zone technique.
$\kappa$ of these crystals was measured as a function of temperature $T$. We used a standard steady state method on pieces cut along the principal axes with a typical length of 2$\,\rm mm$ along the measuring direction and of about 0.5$\,\rm mm$ for the two other directions. The thermal gradient was determined by measuring the temperature difference $\Delta T$ between the
junctions of a differential Au/Fe-Chromel thermocouple. The junctions of this thermocouple have been glued
onto the sample using GE varnish.\cite{foot1} $\Delta T$ varied between 0.5\% and 2\% of the absolute temperature, which has been stabilized for each data point. Errors due to radiation loss, which could occur at higher temperatures, are avoided in our experimental setup.
Stoichiometric oxygen contents in $\rm La_{1.8}Eu_{0.2}CuO_{4}$ and $\rm La_2NiO_4$ were achieved by annealing in high vacuum and $\rm CO/CO_2$-atmosphere, respectively. 

\section{\label{lacuo}Previous results}

\begin{figure}
\includegraphics [width=\columnwidth,clip] {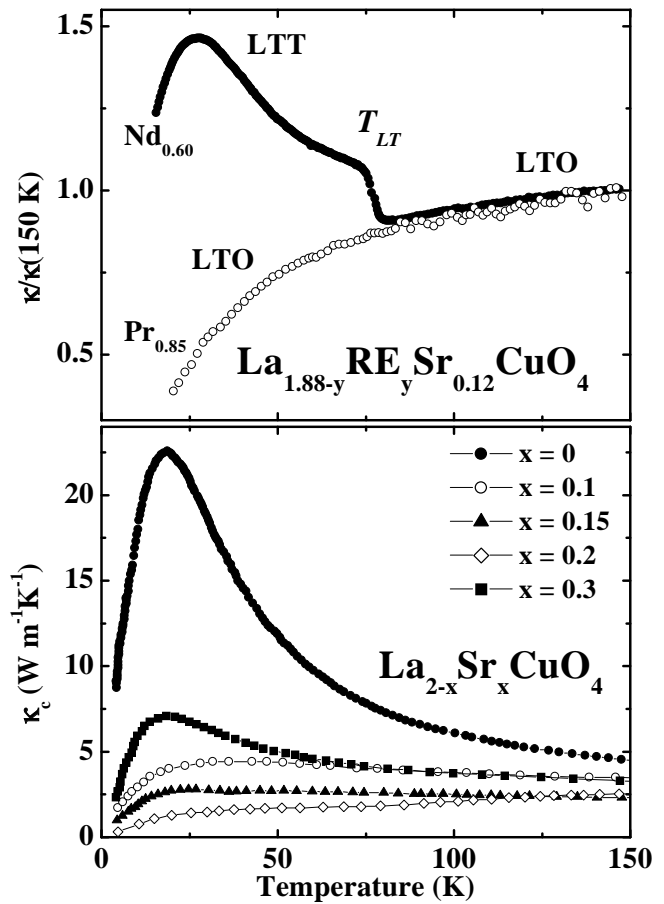}
\caption{\label{fig1}Bottom: Temperature dependence of $\kappa_c$ of $\rm La_{2-x}Sr_xCuO_4$ for
$x=0$, 0.1, 0.15, 0.2 and 0.3. Data are reproduced from Ref.~19. Top: Thermal
conductivity of Pr ($y=0.85$) and Nd ($y=0.6$) doped $\rm La_{1.88-y}(RE)_ySr_{0.12}CuO_4$
polycrystals normalized to $\kappa\rm (150~K)$ as a function of temperature. Data are taken from
Ref.~20.}
\end{figure}
Prior to discussing our new experimental results, we review previous
results\cite{Nakamura91,Baberski98} on the striking doping dependence of the thermal conductivity
of doped $\rm La_2CuO_4$. At first, we concentrate on the low temperature phononic peak in the
thermal conductivity of $\rm La_{2-x}Sr_xCuO_4$. As is evident from the lower panel of
Fig.~\ref{fig1}, which reproduces $\kappa_c$ of $\rm La_{2-x}Sr_xCuO_4$ ($x=0$, 0.1, 0.15, 0.2,
0.3) as published by Nakamura et al.,\cite{Nakamura91} this peak evolves non-monotonically with
increasing Sr-content: A well pronounced phononic low-$T$ peak in $\kappa_c$ is only present at
$x=0$ and $x=0.3$, whereas at intermediate doping levels ($0.1\leq x\leq0.2$) a peak structure is
hardly identifyable. Note that the material is a superconductor in this doping range whereas it is insulating and metallic at $x=0$ and $x=0.3$, respectively.

Baberski et al. have pointed out that the suppressed low-$T$ peak at intermediate Sr-doping
reappears upon RE-doping, provided that this doping induces the so-called LTT-phase (Low
Temperature Tetragonal) and thereby suppresses superconductivity.\cite{Baberski98} This is
illustrated in the upper panel of Fig.~\ref{fig1}, where $\kappa$ of polycrystalline Pr- and Nd-
doped $\rm La_{2-x-y}(RE)_ySr_xCuO_4$ at the finite Sr-content $x=0.12$ is shown. The Pr-doped
compound does not undergo the transition to the LTT-phase. Like $\rm La_{2-x}Sr_xCuO_4$, it remains
in the so-called LTO-phase (Low Temperature Orthorhombic) and is a superconductor.\cite{Schafer94}
Its thermal conductivity monotonically decreases with decreasing $T$ and hence is very similar to
the aforementioned findings by Nakamura et al. for $\kappa$ of $\rm La_{2-x}Sr_xCuO_4$ at $0.1\leq x\leq0.2$. The situation is completely different in the non
superconducting, Nd-doped compound, which is in the LTT-phase below $T_{LT}\approx80$~K: Here,
$\kappa_{\mathrm{ph}}$ abruptly enhances at $T_{LT}$ and exhibits a well pronounced phononic peak
around 25~K.

The doping dependence of $\kappa$ described above must be attributed to strong changes in the
phonon thermal conductivity $\kappa_{\mathrm{ph}}$ of this material: Non-phononic contributions to
$\kappa$ can be excluded in the out-of-plane direction, since the electrical
conductivity\cite{Nakamura92,Nakamura93} and the magnetic coupling\cite{Thio88} along the $c$-axis
are too small to allow significant electronic and magnetic thermal conduction. Concomitantly,
the doping dependence of the low-$T$ peak is evident along all crystal directions\cite{Nakamura91}
(c.f. also the data below).

In crystalline materials, the low-$T$ peak in $\kappa_{\mathrm{ph}}$ is very sensitive to
scattering of phonons. Generally, the height of this peak reduces with increasing phonon scattering
rates. One important mechanism in this regard is scattering of phonons by impurities. In alloyed
compounds like doped $\rm La_2CuO_4$ this phonon-impurity scattering is induced by non-uniform ions
on one lattice site. Upon alloying, this should lead to a gradual reduction of the phonon
peak.\cite{Berman65} Even though phonon-defect scattering inevitably must be present in this
material, it is not a scattering mechanism which solely dominates
$\kappa_{\mathrm{ph}}$, because the phonon peak evolves non-monotonically upon Sr-doping and even
reappears at additional RE-doping.

Another scattering mechanism for phonons in this material is suggested by the abrupt change of $\kappa_{\mathrm{ph}}$ at the structural phase transition, which
occurs in the Nd-doped compound. It appears that phonons are scattered stronger in the
LTO-phase. This interpretation is confirmed by neutron
scattering studies on Nd-doped $\rm La_{2-x}Sr_xCuO_4$ at $T_{LT}$: The acoustic phonon line width
abruptly decreases at the transition from the LTO- into the LTT-phase, signaling a proportional
decrease of scattering processes.\cite{Pintschovius98} Anomalous phonon thermal conductivity in
the vicinity of structural phase transitions is well known from a number of perovskite oxides
as, for example, $\rm SrTiO_3$ and $\rm KTaO_3$. There, a suppression of
$\kappa_{\mathrm{ph}}$ is caused by enhanced scattering of acoustic phonon modes due to their
energetic degeneracy with soft optical phonon modes.\cite{Steigmeier68,Barret70} Indeed, soft
phonon branches do exist in the LTO-phase of doped $\rm
La_2CuO_4$,\cite{Birgeneau87,Pintschovius90,Martinez91,Keimer93} which could cause a suppression of
$\kappa_{\mathrm{ph}}$ (in the following this scattering mechanism will briefly be named
'soft-phonon scattering').
The change of $\kappa_{\mathrm{ph}}$ at $T_{LT}$ then would follow from the
discontinuous hardening of the soft phonon branch in the LTT-phase\cite{Martinez91,Keimer93} and an
associated reduced scattering rate of acoustic phonons.\cite{foot2}

Since all compounds with a suppressed phonon peak are in the LTO-phase at low $T$, soft-phonon
scattering could be important for understanding the suppression of the peak as well.\cite{foot2a} There is,
however, not a one to one correlation between the LTO-phase, i.e., the possible presence of soft phonon
scattering, and the suppression of the peak. This is most obvious in $\rm La_{2-x}Sr_xCuO_4$,
because the soft phonon properties only slightly change for $x\leq 0.2$, whereas the phonon
peak is maximum for $\rm La_2CuO_4$.

Baberski et al. have noticed that both impurity scattering and soft-phonon scattering separately
do not allow to understand the suppression of the phonon peak. They therefore suggested, that the
suppression of the phononic peak could be correlated with the superconducting properties, since the
peak suppression occurs whenever the material is superconducting.\cite{Baberski98} In their model,
they proposed that stripes couple to phonons and thereby cause an unconventional scattering
mechanism for phonons depending on the dynamics of stripes: A suppression of the phonon peak then
is the consequence of fluctuating stripes, which are present in the superconducting compounds,
while $\kappa_{\mathrm{ph}}$ exhibits a usual phonon peak when stripes are static or even absent
in the non-superconducting cases. There, static stripes are only present in the LTT-phase. Therefore, instead of being caused by an abrupt softening of optical phonon modes, the jump of $\kappa_{\mathrm{ph}}$ at $T_{LT}$ could just as well originate from a change of the stripe dynamics from static (LTT) to fluctuating (LTO).
Even though a consistent interpretation of the data discussed afore is possible within such a
model, its validity is questionable because the actual role of impurity- and soft-phonon scattering remains unclear.

\section{\label{data}New results}

In order to elucidate the importance of this conventional scattering mechanisms we now turn
to our measurements on single crystals, which provide profound information in this regard.
\begin{figure}
\includegraphics [width=\columnwidth,clip] {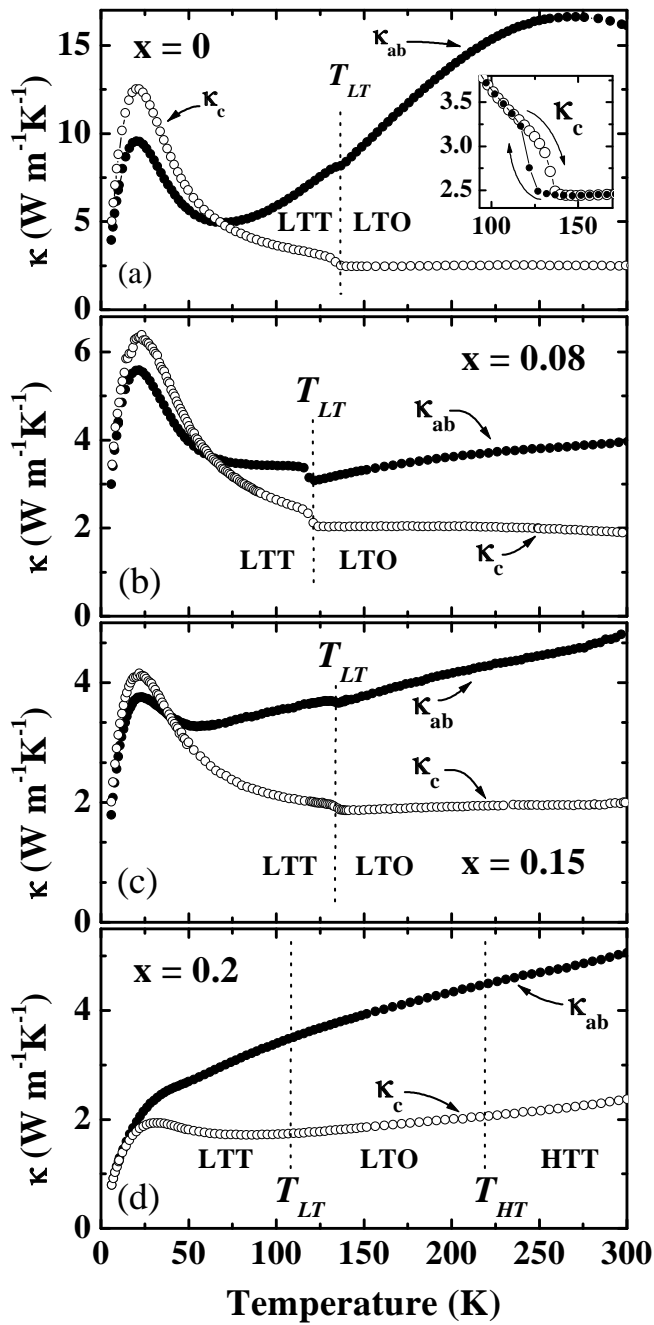}
\caption{\label{fig2}Thermal conductivity of $\rm La_{1.8-x}Eu_{0.2}Sr_xCuO_4$ [$x=0$, 0.08, 0.15, 0.2, panels (a) to (d)] along the $c$-axis ($\kappa_c$, open circles) and parallel to the ab-planes ($\kappa_{ab}$, closed circles) as a function of temperature. Inset: $\kappa_c$ for $x=0$ near $T_{LT}$ at cooling (full circles) and heating (open circles).}
\end{figure}
In Fig.~\ref{fig2} we present the complete data set of $\kappa$ of $\rm
La_{1.8-x}Eu_{0.2}Sr_xCuO_{4}$ ($x=0$, 0.08, 0.15, 0.2) measured along the $c$-axis ($\kappa_{c}$)
and parallel to the $ab$-planes ($\kappa_{ab}$) in the temperature-range $7-300$~K. All compounds
are in the LTT-phase at $T< T_{LT}$, with a variation of $T_{LT}$ between about 110~K and 130~K. For
$T>T_{LT}$ and $x\leq0.15$ the structure is LTO in the investigated temperature range. At $x=0.2$
the compound undergoes a further structural phase transition from the LTO- into the so called
HTT-phase (High-Temperature-Tetragonal) at $T_{HT}\approx220$~K.\cite{Klauss00}

$\kappa_c$ of $\rm La_{1.8}Eu_{0.2}CuO_{4}$ [cf. panel (a) of Fig.~\ref{fig2}] exhibits a low-temperature peak
around 20~K with a falling edge that continuously extends to $T_{LT}\approx130$~K.
A jumplike decrease occurs at $T_{LT}$, followed by a constant $\kappa_c$ up to room
temperature. The inset of Fig.~\ref{fig2} depicts a cooling- and heating curve of $\kappa_c$ in the
area of the jump. A clear hysteresis being a characteristic feature of first-order phase
transitions is evident.

For $T\lesssim50$~K, the thermal conductivity along the $\rm CuO_2$-planes, $\kappa_{ab}$, is comparable to $\kappa_c$. The peak centered around 20~K is slightly smaller; we find
$\kappa_c^{max}/\kappa_{ab}^{max}\approx1.3$, which is similar to the findings in non-doped $\rm
La_2CuO_4$ and isostructural $\rm La_{5/3}Sr_{1/3}NiO_4$.\cite{Nakamura91,Hess99} For
$T\gtrsim50$~K the temperature dependencies of $\kappa_{ab}$ and $\kappa_c$ differ completely: With
rising temperature, $\kappa_{ab}$ strongly increases and evolves into a broad peak at room temperature. A
step-like anomaly is present at $T_{LT}$, which is in a similar way hysteretic as the anomaly in
$\kappa_c$ (not shown).

The panels (b) and (c) of Fig.~\ref{fig2} show $\kappa_{ab}$ and $\kappa_c$ of $\rm
La_{1.8-x}Eu_{0.2}Sr_xCuO_{4}$ for $x=0.08$ and $x=0.15$, respectively. For both compounds
$\kappa_{ab}$ and $\kappa_c$ exhibit a low temperature peak, where the peak of $\kappa_{ab}$ is again slightly smaller than that of $\kappa_c$. The temperature dependence of $\kappa_c$ is very similar to that of $\kappa_c$ of $\rm La_{1.8}Eu_{0.2}CuO_{4}$ in the whole temperature range. As in the case at $x=0$, $\kappa_{ab}$ deviates from the qualitative
$T$-dependence of $\kappa_c$ above $\sim50$~K and evolves into an increase with rising temperature. A steplike
anomaly at $T_{LT}$ exists in each case, but no high temperature peak as in $\rm
La_{1.8}Eu_{0.2}CuO_{4}$ is present.

A comparable anisotropy as in the previous compounds is also evident at $x=0.2$ [cf.
panel (d) of Fig.~\ref{fig2}], but the low temperature peak in $\kappa_c$ is almost completely
suppressed and $\kappa_c$ slightly increases for $T\gtrsim75$~K with nearly constant positive slope.
The low temperature peak is even absent in $\kappa_{ab}$. Here, $\kappa_{ab}(T)$ monotonically
increases upon heating in the entire temperature range. At $T_{LT}\approx110$~K no anomaly is
present, neither in $\kappa_{ab}$ nor in $\kappa_c$. Apparently, the transition at $T_{HT}$ also causes no anomaly in the
thermal conductivity.

%The situation is similar for the low-$T$ peak of $\kappa_{ab}$ (not shown).  $\kappa_{ab}$ is not purely phononic, since, in addition to $\kappa_{\mathrm{ph}}$, magnetic ($x=0$) and electronic ($x>0$) heat conduction are relevant as well. These contributions are, however, rather small at low $T$ and may therefore be ignored in the discussion of the low $T$-peak.

\section{\label{discussion}Discussion}
\subsection{Anisotropy}
\begin{figure}
\includegraphics [width=\columnwidth,clip] {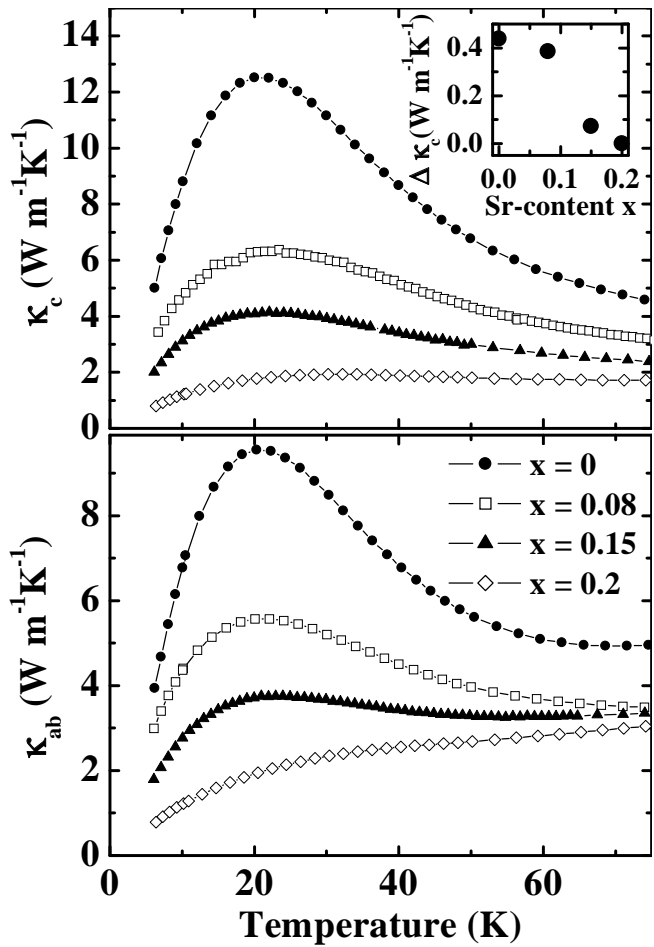}
\caption{\label{fig3}Doping dependence of low temperature peak in the thermal conductivity of $\rm
La_{1.8-x}Eu_{0.2}Sr_xCuO_4$ along the $c$-axis ($\kappa_c$, top) and parallel to the ab-planes
($\kappa_{ab}$, bottom). Inset: Doping dependence of the jump size $\Delta\kappa_c$ at $T_{LT}$.}
\end{figure}

%%%%%%%%%%%%%%%%%%%%%%%%%%%%%%%%%%%%%%%%%%%%%%%FIRST ANALYSIS%%%%%%%%%%%%%%%%%%%%%%%%%%%%%%%%%%%%%%%%%%%%%%%%%%%%%
As mentioned before, $\kappa_c$ is purely phononic, since electronic and magnetic contributions can
be ruled out for heat transport along the $c$-axis. In order to understand the anisotropies of
$\kappa$ for $T\gtrsim50$~K, it is reasonable to distinguish the cases $x=0$ and $x>0$. It has been
shown in Ref.~18 that in insulating, antiferromagnetic $\rm
La_{1.8}Eu_{0.2}CuO_{4}$ the high temperature peak of $\kappa_{ab}$ must be explained by magnetic
contributions. Upon Sr-doping, the doped holes destroy the magnetic order which leads to a strong
suppression of the high temperature peak (cf. Fig.~\ref{fig2}). Simultaneously, the material
becomes electrically conducting within the $ab$-planes.\cite{Hess03,Nakamura91} Therefore, the much
smaller, but apparently still existing anisotropy in the Sr-doped compounds is due to electronic,
rather than magnetic, contributions to $\kappa_{ab}$. Indeed, an estimation using the
Wiedemann-Franz-law yields electronic contributions of the same order of magnitude as the observed
anisotropy for $\kappa_{ab}$ and negligible electronic contributions for $\kappa_c$. The magnetic
and electronic contributions to $\kappa_{ab}$ become important for $T\gtrsim50$~K. Below this
temperature, $\kappa_{ab}$ can be considered to be primarily phononic with a similar magnitude of
$\kappa_{\mathrm{ph}}$ as $\kappa_c$.\cite{foot3} This is nicely confirmed by the very similar
temperature dependence of $\kappa_{ab}$ and $\kappa_c$ in this temperature range, as shown in
Fig.~\ref{fig3}. It is remarkable that for all compounds the phonon peak of $\kappa_c$ is slightly
larger than that of $\kappa_{ab}$ as long as a peak is clearly resolved in both quantities. A
possible explanation could be related to slightly different velocities of the acoustic phonons along the
$ab$- and $c$-directions.\cite{Pintschovius91}

%%%%%%%%%%%%%%%%%%%%%%%%%%%%%%%%%%%%%%%%%%%%%%%%%KAPPA PHONON DISCUSSION%%%%%%%%%%%%%%%%%%%%%%%%%%%%%%%%%%%%%%%%%%
\subsection{Phononic peak and jump at $T_{LT}$}
The development of the low temperature peak of $\kappa_{\mathrm{ph}}$ upon doping can be viewed in
Fig.~\ref{fig3}. Obviously, the peak size is gradually reduced as the Sr-content increases. A
similar result is found for the size of the jump $\Delta\kappa_c$ at $T_{LT}$ which is shown as a
function of doping in the inset of Fig.~\ref{fig3}. 
While the gradual reduction of the peak size can straightforwardly be explained by phonon-impurity scattering, the reduction of the jump at $T_{LT}$ requires further comments. First of all, we stress the presence of a jump in insulating $\rm La_{1.8}Eu_{0.2}CuO_4$. It clearly proofs that it is caused by soft-phonon scattering in the LTO-phase, which therefore has to be regarded as an important scattering channel for phonons indeed. When this scattering channel becomes active in the LTO-phase, the relative importance of phonon-impurity scattering is reduced. The gradual reduction of the jump size
$\Delta\kappa_c$ with increasing Sr-content may therefore be attributed to doping induced phonon-impurity scattering as well. However, we should note that the structural differences between the LTT and LTO-phases diminish with increasing Sr-content. Therefore, apart from phonon-impurity scattering, structural reasons play a role in the suppression of the jump at $T_{LT}$. This is consistent with the observation by Baberski et al., that the jump disappears, whenever the tilting angle of the $\rm CuO_6$-octahedra becomes lower than a critical value.\cite{Baberski98}

%, where a complete suppression of the jump occurs, has been reported to be correlated with the tilting angle of the $\rm CuO_6$-octahedra.\cite{Baberski98} This indicates that apart from phonon-impurity scattering structural peculiarities play a role in the suppression of the jump at $T_{LT}$.\footnote{Since with decreasing tilt angle the differences between the LTT and LTO-phases are reduced, the corresponding changes of the thermal conductivity at $T_{LT}$ could simply be too small to be resolved in the presence of other scattering mechanisms like
%%phonon-impurity scattering, phonon-electron scattering and umklapp scattering, which is already
%%quite effective in this temperature range.}
%
%
%We note, however, that the Sr-doping level, where a complete suppression of the jump occurs, has been reported to be correlated with the tilting angle of the $\rm CuO_6$-octahedra.\cite{Baberski98} This indicates that apart from phonon-impurity scattering structural peculiarities play a role in the suppression of the jump at $T_{LT}$.\footnote{Since with decreasing tilt angle the differences between the LTT and LTO-phases are reduced, the corresponding changes of the thermal conductivity at $T_{LT}$ could simply be too small to be resolved in the presence of other scattering mechanisms like
%phonon-impurity scattering, phonon-electron scattering and umklapp scattering, which is already
%quite effective in this temperature range.}

It is now very instructive to consider $\kappa_{\mathrm{ph}}$ of pure $\rm La_{2-x}Sr_xCuO_4$ (cf. Fig.~\ref{fig1}) for comparison. Apparently, doping Sr into $\rm La_2CuO_4$
suppresses the phonon peak much more effectively than in the Eu-doped counterpart: On the one hand
the phonon peak of non-doped $\rm La_2CuO_4$ is by a factor of about two larger than in non-doped
$\rm La_{1.8}Eu_{0.2}CuO_4$. On the other hand the phonon peaks in $\rm
La_{1.8-x}Eu_{0.2}Sr_xCuO_4$ are clearly better developed for finite Sr-contents with higher
maximum values as in $\rm La_{2-x}Sr_xCuO_4$. One has therefore to conclude that a further scattering mechanism exists in $\rm La_{2-x}Sr_xCuO_4$ which becomes more important upon doping and which is absent or at least much weaker in
the Eu-doped compounds. In this case, this mechanism must cause a more effective scattering than the
surely present phonon-impurity scattering induced by the Eu-ions. One plausible candidate (besides
scattering due to dynamic stripes) for such a mechanism is soft-phonon scattering connected with
the structural instability of the LTO-phase since the soft-phonon energies decrease further upon Sr-doping\cite{foot4} and hence enhance soft-phonon scattering.

%%%%%%%%%%%%%%%%%%%%%%%%%%%%%%%%%%%%%%%%%%%%%%%%%%%%%%%LANIO%%%%%%%%%%%%%%%%%%%%%%%%%%%%%%%%%%%%%%%%%%%%%%%%%%%%
In order to judge the relevance of soft-phonon scattering a measure for its strength with respect
to phonon-impurity scattering is necessary. Such a measure could be achieved, for example, by
comparison with a compound where neither a lattice instability nor doped impurities are present.
Though being no cuprate, isostructural $\rm La_2NiO_4$ is yet a well suited candidate perfectly
fulfilling these requirements. The phonon spectra of $\rm La_2CuO_4$ and $\rm La_2NiO_4$ are almost
identical\cite{Pintschovius89,Pintschovius01} because the atomic masses of Ni and Cu are very
similar. At low temperatures ($T\lesssim73$~K) this compound is in the LTT-phase and hence a
structural instability does not exist.\cite{Rodriguez88} Moreover, no phonon-impurity scattering is
present in this non-doped compound.
\begin{figure}
\includegraphics [width=\columnwidth,clip] {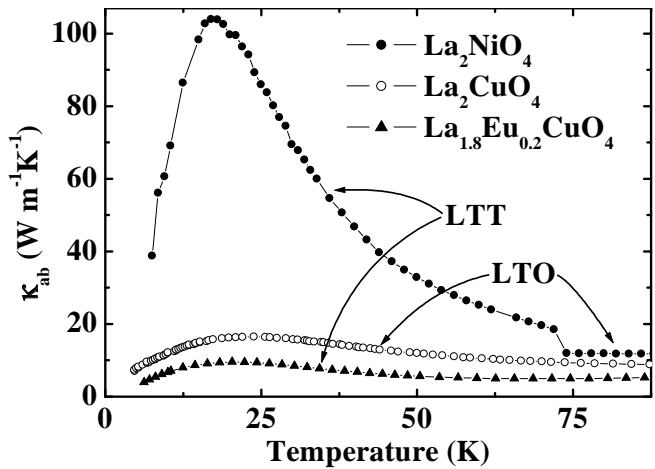}
\caption{\label{fig4}Temperature dependence of $\kappa_{ab}$ of $\rm La_2NiO_4$ in comparison with
$\kappa_{ab}$ of $\rm La_2CuO_4$ and $\rm La_{1.8}Eu_{0.2}CuO_4$. The data for $\rm La_2CuO_4$ are
reproduced from Ref.~\cite{Nakamura91}.}
\end{figure}

Fig.~\ref{fig4} presents our result for phononic\cite{foot5} $\kappa_{ab}$ of
electrically insulating $\rm La_2NiO_4$ in comparison with $\kappa_{ab}$ of $\rm La_2CuO_4$ as
measured by Nakamura et al. and with $\kappa_{ab}$ of $\rm La_{1.8}Eu_{0.2}CuO_4$. It is intriguing
that the phonon peak of $\rm La_2NiO_4$ is about one order of magnitude larger than the peak in
both $\rm La_2CuO_4$ and $\rm La_{1.8}Eu_{0.2}CuO_4$. At $T_{LT}\approx73$~K, the structure changes
from LTT to LTO, which also in this material causes a jumplike decrease in the thermal
conductivity. Compared to $\rm La_{1.8}Eu_{0.2}CuO_4$ the jump size is larger by a factor of about
10. In the LTO-phase, $\kappa_{ab}$ is of the similar size as in $\rm La_2CuO_4$.

It immediately follows from this observation that in $\rm La_2CuO_4$ as well as in $\rm
La_{1.8}Eu_{0.2}CuO_4$ the heat carrying phonons are subject of severe scattering at low
temperatures. In both cases this scattering obviously suppresses the peak of $\kappa_{\mathrm{ph}}$
by about one order of magnitude. Translated to the strength of the two discussed scattering
mechanisms this means that referring to a non-impurity-doped LTT-phase both soft-phonon scattering
and phonon-impurity scattering independently of each other reduce the peak of
$\kappa_{\mathrm{ph}}$ by almost the same amount.

This conclusion now allows us to understand the doping dependence of the phonon peak, when a
combination of both, phonon-impurity- and soft-phonon scattering is consired: In $\rm
La_{1.8-x}Eu_{0.2}Sr_xCuO_4$, where the thermal conductivity is already strongly influenced by the
Eu-ions, Sr-doping simply further increases the phonon-impurity scattering rate. This consequently
leads to a gradual reduction of the phonon peak with increasing Sr-content. Soft-phonon scattering
is not relevant here.

In $\rm La_{2-x}Sr_xCuO_4$ with the same Sr-content than a Eu-doped counterpart, the
phonon-impurity scattering rate is slightly reduced since no Eu-ions are present. Yet, the
resulting phonon thermal conductivity is somewhat smaller. Hence, the effect of a slightly reduced
phonon-impurity scattering rate must be overcompensated by soft-phonon scattering. This is indeed reasonable since it is a qualitatively different scattering mechanism whose importance upon Sr-doping grows since thereby the relevant soft modes soften further.\cite{Braden94}

The reappearance of the phonon peak in $\rm La_{2-x}Sr_xCuO_4$ at high Sr-concentrations as $x=0.3$
can now be understood as a natural consequence of the doping dependence of $T_{HT}$. For $x\gtrsim0.22$,
the LTO-phase disappears in $\rm La_{2-x}Sr_xCuO_4$ and the structure remains in the HTT-phase for
all temperatures.\cite{Takagi92a} Hence, for $x=0.3$ soft-phonon scattering associated with the
LTO-phase is not active for such overdoped compounds. The result of such decreased phonon
scattering is the reappearance of the phonon peak.

It is necessary to mention that the growing density of holes as the Sr-content is increased in
principle could be a further source of scattering and therefore contribute to the reduction of
$\kappa_{\mathrm{ph}}$ and in particular the phonon peak via phonon-electron scattering. However,
since the transition to superconductivity in $\rm La_{2-x}Sr_xCuO_4$ causes no significant anomaly
in $\kappa_c$ at $T_c$ (cf. Fig~\ref{fig1}), this scattering mechanism is usually considered to be
unimportant in this material and its (RE)-doped relatives.\cite{Baberski98} Yet if this scattering
channel is active in this material no inconsistency to our interpretation arises because in this
case the phonon-'impurity' scattering induced by Sr-ions can be regarded simply as slightly more
effective than the phonon-impurity scattering induced by Eu-ions.

\subsection{Stripes as a scattering mechanism for phonons?}
The above discussion of $\kappa_{\mathrm{ph}}$ of the insulating materials $\rm La_{2}CuO_4$, $\rm La_{1.8}Eu_{0.2}CuO_4$ and $\rm La_2NiO_4$ provides unambiguous evidence that phonon-impurity and soft-phonon scattering are important scattering mechanisms for phonons in doped $\rm La_{2}CuO_4$. Since the major doping dependencies of the low-$T$ peak can be explained based on these mechanisms without any problems, there is no compelling reason to incorporate a stripe induced scattering channel in the data interpretation. It is unlikely though that stripes in doped $\rm La_{2}CuO_4$ have no effect at all on $\kappa_{\mathrm{ph}}$ because the aforementioned stripe-induced phonon scattering in the nickelates\cite{Hess99,Cassel99a} is an unambiguous physical fact. However, it appears extremely difficult to prove the existence and to study the strength of purely stripe induced scattering in doped $\rm La_{2}CuO_4$ via $\kappa_{\mathrm{ph}}$. This is not only because phonon-impurity and soft-phonon scattering are already dominating $\kappa_{\mathrm{ph}}$ in insulating $\rm La_{2}CuO_4$; Due to the intimate relation of LTT-phase and stripe dynamics, scattering on fluctuating stripes in the Sr-doped compounds could be viewed as an altered but already existing soft-phonon scattering, i.e., from this point of view these two scattering mechanisms are conceptually indistinguishable.

\subsection{High temperature increase of $\kappa_{\mathrm{ph}}$}
For completeness, we briefly mention the unusual temperature dependence of $\kappa_c$ which is evident at higher temperatures $T\gtrsim100$~K. Instead of a decrease as $\sim T^{-1}$, which at elevated temperatures is usually expected for thermal conductivity by {\em acoustic} phonons, $\kappa_{c}$ is almost temperature independent or even increases with rising $T$. As has been shown in Ref.~60, this strong deviation from the usual behavior arises due to thermal conduction by dispersive {\em optical} phonons in addition to the usual contribution by acoustic phonons.

\section{Summary}
%%%%%%%%%%%%%%%%%%%%%%%%%%%%%%%%%%%%%%%%%%%SUMMARY%%%%%%%%%%%%%%%%%%%%%%%%%%%%%%%%%%%%%%%%%%%%%%%%%%%%%%%%%%%%%%%%%%%%%
In summary, we have presented experimental results on the thermal conductivity of $\rm La_{1.8-x}Eu_{0.2}Sr_xCuO_4$ for a wide doping range of Sr. The analysis of our data suggests that in this material phonons are strongly scattered on the doped impurities, i.e., Sr and Eu, as well as on soft phonons that are present in the LTO-phase of this material.
Comparison of our data with the thermal conductivity of $\rm La_{2-x}Sr_xCuO_4$ and isostructural $\rm La_2NiO_4$ leads to the conclusion that these scattering mechanisms are most relevant in $\rm La_{2-x}Sr_xCuO_4$ as well. In particular, in contrast to other studies\cite{Baberski98,Sun03,Sun03a}, there remains no direct evidence that the stripe correlations cause a relevant scattering channel in doped $\rm La_2CuO_4$.

\section{Acknowledgments}
%%%%%%%%%%%%%%%%%%%%%%%%%%%%%%%%%%%%%%%ACKNOWLEDGMENTS%%%%%%%%%%%%%%%%%%%%%%%%%%%%%%%%%%%%%%%%
C.H. acknowledges support by the DFG through HE3439/3-1. We thank N. Jenkins for proofreading the manuscript.

%\bibliography{lacuoph}

\begin{thebibliography}{60}
\expandafter\ifx\csname natexlab\endcsname\relax\def\natexlab#1{#1}\fi
\expandafter\ifx\csname bibnamefont\endcsname\relax
  \def\bibnamefont#1{#1}\fi
\expandafter\ifx\csname bibfnamefont\endcsname\relax
  \def\bibfnamefont#1{#1}\fi
\expandafter\ifx\csname citenamefont\endcsname\relax
  \def\citenamefont#1{#1}\fi
\expandafter\ifx\csname url\endcsname\relax
  \def\url#1{\texttt{#1}}\fi
\expandafter\ifx\csname urlprefix\endcsname\relax\def\urlprefix{URL }\fi
\providecommand{\bibinfo}[2]{#2}
\providecommand{\eprint}[2][]{\url{#2}}

\bibitem[{\citenamefont{Tranquada et~al.}(1994)\citenamefont{Tranquada,
  Buttrey, Sachan, and Lorenzo}}]{Tranquada94}
\bibinfo{author}{\bibfnamefont{J.~M.} \bibnamefont{Tranquada}},
  \bibinfo{author}{\bibfnamefont{D.~J.} \bibnamefont{Buttrey}},
  \bibinfo{author}{\bibfnamefont{V.}~\bibnamefont{Sachan}}, \bibnamefont{and}
  \bibinfo{author}{\bibfnamefont{J.~E.} \bibnamefont{Lorenzo}},
  \bibinfo{journal}{Phys. Rev. Lett.} \textbf{\bibinfo{volume}{73}},
  \bibinfo{pages}{1003} (\bibinfo{year}{1994}).

\bibitem[{\citenamefont{Tranquada
  et~al.}(1995{\natexlab{a}})\citenamefont{Tranquada, Sternlieb, Axe, Nakamura,
  and Uchida}}]{Tranquada95}
\bibinfo{author}{\bibfnamefont{J.~M.} \bibnamefont{Tranquada}},
  \bibinfo{author}{\bibfnamefont{B.~J.} \bibnamefont{Sternlieb}},
  \bibinfo{author}{\bibfnamefont{J.~D.} \bibnamefont{Axe}},
  \bibinfo{author}{\bibfnamefont{Y.}~\bibnamefont{Nakamura}}, \bibnamefont{and}
  \bibinfo{author}{\bibfnamefont{S.}~\bibnamefont{Uchida}},
  \bibinfo{journal}{Nature} \textbf{\bibinfo{volume}{375}},
  \bibinfo{pages}{561} (\bibinfo{year}{1995}{\natexlab{a}}).

\bibitem[{\citenamefont{Tranquada
  et~al.}(1995{\natexlab{b}})\citenamefont{Tranquada, Lorenzo, Buttrey, and
  Sachan}}]{Tranquada95a}
\bibinfo{author}{\bibfnamefont{J.~M.} \bibnamefont{Tranquada}},
  \bibinfo{author}{\bibfnamefont{J.~E.} \bibnamefont{Lorenzo}},
  \bibinfo{author}{\bibfnamefont{D.~J.} \bibnamefont{Buttrey}},
  \bibnamefont{and} \bibinfo{author}{\bibfnamefont{V.}~\bibnamefont{Sachan}},
  \bibinfo{journal}{Phys. Rev. B} \textbf{\bibinfo{volume}{52}},
  \bibinfo{pages}{3581} (\bibinfo{year}{1995}{\natexlab{b}}).

\bibitem[{\citenamefont{Lee and Cheong}(1997)}]{Lee97}
\bibinfo{author}{\bibfnamefont{S.-H.} \bibnamefont{Lee}} \bibnamefont{and}
  \bibinfo{author}{\bibfnamefont{S.-W.} \bibnamefont{Cheong}},
  \bibinfo{journal}{Phys. Rev. Lett.} \textbf{\bibinfo{volume}{79}},
  \bibinfo{pages}{2514} (\bibinfo{year}{1997}).

\bibitem[{\citenamefont{Mori et~al.}(1998)\citenamefont{Mori, Chen, and
  Cheong}}]{Mori98}
\bibinfo{author}{\bibfnamefont{S.}~\bibnamefont{Mori}},
  \bibinfo{author}{\bibfnamefont{C.~H.} \bibnamefont{Chen}}, \bibnamefont{and}
  \bibinfo{author}{\bibfnamefont{S.-W.} \bibnamefont{Cheong}},
  \bibinfo{journal}{Nature} \textbf{\bibinfo{volume}{392}},
  \bibinfo{pages}{473} (\bibinfo{year}{1998}).

\bibitem[{\citenamefont{White and Scalapino}(1999)}]{White99}
\bibinfo{author}{\bibfnamefont{S.~R.} \bibnamefont{White}} \bibnamefont{and}
  \bibinfo{author}{\bibfnamefont{D.~J.} \bibnamefont{Scalapino}},
  \bibinfo{journal}{Phys. Rev. B} \textbf{\bibinfo{volume}{60}},
  \bibinfo{pages}{753} (\bibinfo{year}{1999}).

\bibitem[{\citenamefont{Chakravarty et~al.}(2001)\citenamefont{Chakravarty,
  Laughlin, Morr, and Nayak}}]{Chakravarty01}
\bibinfo{author}{\bibfnamefont{S.}~\bibnamefont{Chakravarty}},
  \bibinfo{author}{\bibfnamefont{R.~B.} \bibnamefont{Laughlin}},
  \bibinfo{author}{\bibfnamefont{D.~K.} \bibnamefont{Morr}}, \bibnamefont{and}
  \bibinfo{author}{\bibfnamefont{C.}~\bibnamefont{Nayak}},
  \bibinfo{journal}{Phys. Rev. B} \textbf{\bibinfo{volume}{63}},
  \bibinfo{pages}{094503} (\bibinfo{year}{2001}).

\bibitem[{\citenamefont{Zaanen}(2003)}]{Zaanen03}
\bibinfo{author}{\bibfnamefont{J.}~\bibnamefont{Zaanen}},
  \bibinfo{journal}{Nature} \textbf{\bibinfo{volume}{422}},
  \bibinfo{pages}{569} (\bibinfo{year}{2003}).

\bibitem[{\citenamefont{B{\"u}chner et~al.}(1994)\citenamefont{B\"{u}chner,
  Breuer, Freimuth, and Kampf}}]{Buchner94}
\bibinfo{author}{\bibfnamefont{B.}~\bibnamefont{B\"{u}chner}},
  \bibinfo{author}{\bibfnamefont{M.}~\bibnamefont{Breuer}},
  \bibinfo{author}{\bibfnamefont{A.}~\bibnamefont{Freimuth}}, \bibnamefont{and}
  \bibinfo{author}{\bibfnamefont{A.~P.} \bibnamefont{Kampf}},
  \bibinfo{journal}{Phys. Rev. Lett.} \textbf{\bibinfo{volume}{73}},
  \bibinfo{pages}{1841} (\bibinfo{year}{1994}).

\bibitem[{\citenamefont{Klauss et~al.}(2000)\citenamefont{Klauss, Wagener,
  Hillberg, Kopmann, Walf, Litterst, H\"{u}cker, and B\"{u}chner}}]{Klauss00}
\bibinfo{author}{\bibfnamefont{H.-H.} \bibnamefont{Klauss}},
  \bibinfo{author}{\bibfnamefont{W.}~\bibnamefont{Wagener}},
  \bibinfo{author}{\bibfnamefont{M.}~\bibnamefont{Hillberg}},
  \bibinfo{author}{\bibfnamefont{W.}~\bibnamefont{Kopmann}},
  \bibinfo{author}{\bibfnamefont{H.}~\bibnamefont{Walf}},
  \bibinfo{author}{\bibfnamefont{F.~J.} \bibnamefont{Litterst}},
  \bibinfo{author}{\bibfnamefont{M.}~\bibnamefont{H\"{u}cker}},
  \bibnamefont{and}
  \bibinfo{author}{\bibfnamefont{B.}~\bibnamefont{B\"{u}chner}},
  \bibinfo{journal}{Phys. Rev. Lett.} \textbf{\bibinfo{volume}{85}},
  \bibinfo{pages}{4590} (\bibinfo{year}{2000}).

\bibitem[{\citenamefont{Lake et~al.}(2001)\citenamefont{Lake, Aeppli, Clausen,
  McMorrow, Lefmann, Hussey, Mangkorntong, Nohara, Takagi, Mason
  et~al.}}]{Lake01}
\bibinfo{author}{\bibfnamefont{B.}~\bibnamefont{Lake}},
  \bibinfo{author}{\bibfnamefont{G.}~\bibnamefont{Aeppli}},
  \bibinfo{author}{\bibfnamefont{K.}~\bibnamefont{Clausen}},
  \bibinfo{author}{\bibfnamefont{D.}~\bibnamefont{McMorrow}},
  \bibinfo{author}{\bibfnamefont{K.}~\bibnamefont{Lefmann}},
  \bibinfo{author}{\bibfnamefont{N.~E.} \bibnamefont{Hussey}},
  \bibinfo{author}{\bibfnamefont{N.}~\bibnamefont{Mangkorntong}},
  \bibinfo{author}{\bibfnamefont{M.}~\bibnamefont{Nohara}},
  \bibinfo{author}{\bibfnamefont{H.}~\bibnamefont{Takagi}},
  \bibinfo{author}{\bibfnamefont{T.~E.} \bibnamefont{Mason}},
  \bibnamefont{et~al.}, \bibinfo{journal}{Science}
  \textbf{\bibinfo{volume}{291}}, \bibinfo{pages}{1759} (\bibinfo{year}{2001}).

\bibitem[{\citenamefont{Lake et~al.}(2002)\citenamefont{Lake, R{\o}nnow,
  Christensen, Aeppli, Lefmann, McMorrow, Vorderwisch, Smeibidl, Mangkorntong,
  Sasagawa et~al.}}]{Lake02}
\bibinfo{author}{\bibfnamefont{B.}~\bibnamefont{Lake}},
  \bibinfo{author}{\bibfnamefont{H.~M.} \bibnamefont{R{\o}nnow}},
  \bibinfo{author}{\bibfnamefont{N.~B.} \bibnamefont{Christensen}},
  \bibinfo{author}{\bibfnamefont{G.}~\bibnamefont{Aeppli}},
  \bibinfo{author}{\bibfnamefont{K.}~\bibnamefont{Lefmann}},
  \bibinfo{author}{\bibfnamefont{D.~F.} \bibnamefont{McMorrow}},
  \bibinfo{author}{\bibfnamefont{P.}~\bibnamefont{Vorderwisch}},
  \bibinfo{author}{\bibfnamefont{P.}~\bibnamefont{Smeibidl}},
  \bibinfo{author}{\bibfnamefont{N.}~\bibnamefont{Mangkorntong}},
  \bibinfo{author}{\bibfnamefont{T.}~\bibnamefont{Sasagawa}},
  \bibnamefont{et~al.}, \bibinfo{journal}{Nature}
  \textbf{\bibinfo{volume}{415}}, \bibinfo{pages}{299} (\bibinfo{year}{2002}).

\bibitem[{\citenamefont{Tranquada et~al.}(1997)\citenamefont{Tranquada, Axe,
  Ichikawa, Moodenbaugh, Nakamura, and Uchida}}]{Tranquada97}
\bibinfo{author}{\bibfnamefont{J.~M.} \bibnamefont{Tranquada}},
  \bibinfo{author}{\bibfnamefont{J.~D.} \bibnamefont{Axe}},
  \bibinfo{author}{\bibfnamefont{N.}~\bibnamefont{Ichikawa}},
  \bibinfo{author}{\bibfnamefont{A.~R.} \bibnamefont{Moodenbaugh}},
  \bibinfo{author}{\bibfnamefont{Y.}~\bibnamefont{Nakamura}}, \bibnamefont{and}
  \bibinfo{author}{\bibfnamefont{S.}~\bibnamefont{Uchida}},
  \bibinfo{journal}{Phys. Rev. Lett.} \textbf{\bibinfo{volume}{78}},
  \bibinfo{pages}{338} (\bibinfo{year}{1997}).

\bibitem[{\citenamefont{Tranquada et~al.}(1999)\citenamefont{Tranquada,
  Ichikawa, and Uchida}}]{Tranquada99}
\bibinfo{author}{\bibfnamefont{J.~M.} \bibnamefont{Tranquada}},
  \bibinfo{author}{\bibfnamefont{N.}~\bibnamefont{Ichikawa}}, \bibnamefont{and}
  \bibinfo{author}{\bibfnamefont{S.}~\bibnamefont{Uchida}},
  \bibinfo{journal}{Phys. Rev. B} \textbf{\bibinfo{volume}{59}},
  \bibinfo{pages}{14712} (\bibinfo{year}{1999}).

\bibitem[{\citenamefont{Cheong et~al.}(1991)\citenamefont{Cheong, Aeppli,
  Mason, Mook, Hayden, Canfield, Fisk, Clausen, and Martinez}}]{Cheong91}
\bibinfo{author}{\bibfnamefont{S.-W.} \bibnamefont{Cheong}},
  \bibinfo{author}{\bibfnamefont{G.}~\bibnamefont{Aeppli}},
  \bibinfo{author}{\bibfnamefont{T.~E.} \bibnamefont{Mason}},
  \bibinfo{author}{\bibfnamefont{H.}~\bibnamefont{Mook}},
  \bibinfo{author}{\bibfnamefont{S.~M.} \bibnamefont{Hayden}},
  \bibinfo{author}{\bibfnamefont{P.~C.} \bibnamefont{Canfield}},
  \bibinfo{author}{\bibfnamefont{Z.}~\bibnamefont{Fisk}},
  \bibinfo{author}{\bibfnamefont{K.~N.} \bibnamefont{Clausen}},
  \bibnamefont{and} \bibinfo{author}{\bibfnamefont{J.~L.}
  \bibnamefont{Martinez}}, \bibinfo{journal}{Phys. Rev. Lett.}
  \textbf{\bibinfo{volume}{67}}, \bibinfo{pages}{1791} (\bibinfo{year}{1991}).

\bibitem[{\citenamefont{Yamada et~al.}(1998)\citenamefont{Yamada, Lee,
  Kurahashi, Wada, Wakimoto, Ueki, Kimura, Endoh, Hosoya, Shirane
  et~al.}}]{Yamada98}
\bibinfo{author}{\bibfnamefont{K.}~\bibnamefont{Yamada}},
  \bibinfo{author}{\bibfnamefont{C.~H.} \bibnamefont{Lee}},
  \bibinfo{author}{\bibfnamefont{K.}~\bibnamefont{Kurahashi}},
  \bibinfo{author}{\bibfnamefont{J.}~\bibnamefont{Wada}},
  \bibinfo{author}{\bibfnamefont{S.}~\bibnamefont{Wakimoto}},
  \bibinfo{author}{\bibfnamefont{S.}~\bibnamefont{Ueki}},
  \bibinfo{author}{\bibfnamefont{H.}~\bibnamefont{Kimura}},
  \bibinfo{author}{\bibfnamefont{Y.}~\bibnamefont{Endoh}},
  \bibinfo{author}{\bibfnamefont{S.}~\bibnamefont{Hosoya}},
  \bibinfo{author}{\bibfnamefont{G.}~\bibnamefont{Shirane}},
  \bibinfo{author}{\bibfnamefont{R.~J.}~\bibnamefont{Birgeneau}},
  \bibinfo{author}{\bibfnamefont{M.}~\bibnamefont{Greven}},
  \bibinfo{author}{\bibfnamefont{M.~A.}~\bibnamefont{Kastner}},
  \bibnamefont{and} \bibinfo{author}{\bibfnamefont{Y.~J.}
  \bibnamefont{Kim}},
  \bibinfo{journal}{Phys. Rev. B}
  \textbf{\bibinfo{volume}{57}}, \bibinfo{pages}{6165} (\bibinfo{year}{1998}).

\bibitem[{\citenamefont{Mook et~al.}(1998)\citenamefont{Mook, Dai, Hayden,
  Aeppli, Perring, and Do\v{g}an}}]{Mook98}
\bibinfo{author}{\bibfnamefont{H.~A.} \bibnamefont{Mook}},
  \bibinfo{author}{\bibfnamefont{P.}~\bibnamefont{Dai}},
  \bibinfo{author}{\bibfnamefont{S.~M.} \bibnamefont{Hayden}},
  \bibinfo{author}{\bibfnamefont{G.}~\bibnamefont{Aeppli}},
  \bibinfo{author}{\bibfnamefont{T.~G.} \bibnamefont{Perring}},
  \bibnamefont{and}
  \bibinfo{author}{\bibfnamefont{F.}~\bibnamefont{Do\v{g}an}},
  \bibinfo{journal}{Nature} \textbf{\bibinfo{volume}{395}},
  \bibinfo{pages}{580} (\bibinfo{year}{1998}).

\bibitem[{\citenamefont{Hess et~al.}(2003)\citenamefont{Hess, B\"{u}chner,
  Ammerahl, Colonescu, Heidrich-Meisner, Brenig, and Revcolevschi}}]{Hess03}
\bibinfo{author}{\bibfnamefont{C.}~\bibnamefont{Hess}},
  \bibinfo{author}{\bibfnamefont{B.}~\bibnamefont{B\"{u}chner}},
  \bibinfo{author}{\bibfnamefont{U.}~\bibnamefont{Ammerahl}},
  \bibinfo{author}{\bibfnamefont{L.}~\bibnamefont{Colonescu}},
  \bibinfo{author}{\bibfnamefont{F.}~\bibnamefont{Heidrich-Meisner}},
  \bibinfo{author}{\bibfnamefont{W.}~\bibnamefont{Brenig}}, \bibnamefont{and}
  \bibinfo{author}{\bibfnamefont{A.}~\bibnamefont{Revcolevschi}},
  \bibinfo{journal}{Phys. Rev. Lett.} \textbf{\bibinfo{volume}{90}},
  \bibinfo{pages}{197002} 
  (\bibinfo{year}{2003}).

\bibitem[{\citenamefont{Nakamura et~al.}(1991)\citenamefont{Nakamura, Uchida,
  Kimura, Motohira, Kishio, Kitazawa, Arima, and Tokura}}]{Nakamura91}
\bibinfo{author}{\bibfnamefont{Y.}~\bibnamefont{Nakamura}},
  \bibinfo{author}{\bibfnamefont{S.}~\bibnamefont{Uchida}},
  \bibinfo{author}{\bibfnamefont{T.}~\bibnamefont{Kimura}},
  \bibinfo{author}{\bibfnamefont{N.}~\bibnamefont{Motohira}},
  \bibinfo{author}{\bibfnamefont{K.}~\bibnamefont{Kishio}},
  \bibinfo{author}{\bibfnamefont{K.}~\bibnamefont{Kitazawa}},
  \bibinfo{author}{\bibfnamefont{T.}~\bibnamefont{Arima}}, \bibnamefont{and}
  \bibinfo{author}{\bibfnamefont{Y.}~\bibnamefont{Tokura}},
  \bibinfo{journal}{Physica C} \textbf{\bibinfo{volume}{185-189}},
  \bibinfo{pages}{1409} (\bibinfo{year}{1991}).

\bibitem[{\citenamefont{Baberski et~al.}(1998)\citenamefont{Baberski, Lang,
  Maldonado, H\"{u}cker, B\"{u}chner, and Freimuth}}]{Baberski98}
\bibinfo{author}{\bibfnamefont{O.}~\bibnamefont{Baberski}},
  \bibinfo{author}{\bibfnamefont{A.}~\bibnamefont{Lang}},
  \bibinfo{author}{\bibfnamefont{O.}~\bibnamefont{Maldonado}},
  \bibinfo{author}{\bibfnamefont{M.}~\bibnamefont{H\"{u}cker}},
  \bibinfo{author}{\bibfnamefont{B.}~\bibnamefont{B\"{u}chner}},
  \bibnamefont{and} \bibinfo{author}{\bibfnamefont{A.}~\bibnamefont{Freimuth}},
  \bibinfo{journal}{Europhys. Lett.} \textbf{\bibinfo{volume}{44}},
  \bibinfo{pages}{335} (\bibinfo{year}{1998}).

\bibitem[{\citenamefont{Sun et~al.}(2003{\natexlab{a}})\citenamefont{Sun,
  Takeya, Komiya, and Ando}}]{Sun03}
\bibinfo{author}{\bibfnamefont{X.~F.} \bibnamefont{Sun}},
  \bibinfo{author}{\bibfnamefont{J.}~\bibnamefont{Takeya}},
  \bibinfo{author}{\bibfnamefont{S.}~\bibnamefont{Komiya}}, \bibnamefont{and}
  \bibinfo{author}{\bibfnamefont{Y.}~\bibnamefont{Ando}},
  \bibinfo{journal}{Phys. Rev. B} \textbf{\bibinfo{volume}{67}},
  \bibinfo{pages}{104503} (\bibinfo{year}{2003}{\natexlab{a}}).

\bibitem[{\citenamefont{Sun et~al.}(2003{\natexlab{b}})\citenamefont{Sun,
  Komiya, and Ando}}]{Sun03a}
\bibinfo{author}{\bibfnamefont{X.~F.} \bibnamefont{Sun}},
  \bibinfo{author}{\bibfnamefont{S.}~\bibnamefont{Komiya}}, \bibnamefont{and}
  \bibinfo{author}{\bibfnamefont{Y.}~\bibnamefont{Ando}},
  \bibinfo{journal}{Phys. Rev. B} \textbf{\bibinfo{volume}{67}},
  \bibinfo{pages}{184512} (\bibinfo{year}{2003}{\natexlab{b}}).

\bibitem[{\citenamefont{Sun et~al.}(2003{\natexlab{b}})\citenamefont{Sun, Kurita, Suzuki, 
  Komiya, and Ando}}]{Sun03b}
\bibinfo{author}{\bibfnamefont{X.~F.} \bibnamefont{Sun}},
\bibinfo{author}{\bibfnamefont{Y.} \bibnamefont{Kurita}},
\bibinfo{author}{\bibfnamefont{T.} \bibnamefont{Suzuki}},
  \bibinfo{author}{\bibfnamefont{S.}~\bibnamefont{Komiya}}, \bibnamefont{and}
  \bibinfo{author}{\bibfnamefont{Y.}~\bibnamefont{Ando}},
  \bibinfo{journal}{cond-mat/0308263 (unpublished)}.


\bibitem[{\citenamefont{Hess et~al.}(1999)\citenamefont{Hess, B\"{u}chner,
  H\"{u}cker, Gross, and Cheong}}]{Hess99}
\bibinfo{author}{\bibfnamefont{C.}~\bibnamefont{Hess}},
  \bibinfo{author}{\bibfnamefont{B.}~\bibnamefont{B\"{u}chner}},
  \bibinfo{author}{\bibfnamefont{M.}~\bibnamefont{H\"{u}cker}},
  \bibinfo{author}{\bibfnamefont{R.}~\bibnamefont{Gross}}, \bibnamefont{and}
  \bibinfo{author}{\bibfnamefont{S.-W.} \bibnamefont{Cheong}},
  \bibinfo{journal}{Phys. Rev. B} \textbf{\bibinfo{volume}{59}},
  \bibinfo{pages}{10397} (\bibinfo{year}{1999}).

\bibitem[{\citenamefont{Cassel et~al.}(1999)\citenamefont{Cassel, Hess,
  B\"{u}chner, H\"{u}cker, Gross, Friedt, and Cheong}}]{Cassel99a}
\bibinfo{author}{\bibfnamefont{D.}~\bibnamefont{Cassel}},
  \bibinfo{author}{\bibfnamefont{C.}~\bibnamefont{Hess}},
  \bibinfo{author}{\bibfnamefont{B.}~\bibnamefont{B\"{u}chner}},
  \bibinfo{author}{\bibfnamefont{M.}~\bibnamefont{H\"{u}cker}},
  \bibinfo{author}{\bibfnamefont{R.}~\bibnamefont{Gross}},
  \bibinfo{author}{\bibfnamefont{O.}~\bibnamefont{Friedt}}, \bibnamefont{and}
  \bibinfo{author}{\bibfnamefont{S.-W.} \bibnamefont{Cheong}},
  \bibinfo{journal}{J. Low Temp. Phys.} \textbf{\bibinfo{volume}{117}},
  \bibinfo{pages}{1083} (\bibinfo{year}{1999}).

\bibitem[{\citenamefont{Lee et~al.}(2002)\citenamefont{Lee, Tranquada, Yamada,
  Buttrey, Li, and Cheong}}]{Lee02}
\bibinfo{author}{\bibfnamefont{S.-H.} \bibnamefont{Lee}},
  \bibinfo{author}{\bibfnamefont{J.~M.} \bibnamefont{Tranquada}},
  \bibinfo{author}{\bibfnamefont{K.}~\bibnamefont{Yamada}},
  \bibinfo{author}{\bibfnamefont{D.~J.} \bibnamefont{Buttrey}},
  \bibinfo{author}{\bibfnamefont{Q.}~\bibnamefont{Li}}, \bibnamefont{and}
  \bibinfo{author}{\bibfnamefont{S.-W.} \bibnamefont{Cheong}},
  \bibinfo{journal}{Phys. Rev. Lett.} \textbf{\bibinfo{volume}{88}},
  \bibinfo{pages}{126401} (\bibinfo{year}{2002}).

\bibitem[{\citenamefont{Eisaki et~al.}(1992)\citenamefont{Eisaki, Uchida,
  Mizokawa, Namatame, Fujimori, van Elp, Kuiper, Sawatzky, Hosoya, and
  Yoshida}}]{Eisaki92}
\bibinfo{author}{\bibfnamefont{H.}~\bibnamefont{Eisaki}},
  \bibinfo{author}{\bibfnamefont{S.}~\bibnamefont{Uchida}},
  \bibinfo{author}{\bibfnamefont{T.}~\bibnamefont{Mizokawa}},
  \bibinfo{author}{\bibfnamefont{H.}~\bibnamefont{Namatame}},
  \bibinfo{author}{\bibfnamefont{A.}~\bibnamefont{Fujimori}},
  \bibinfo{author}{\bibfnamefont{J.}~\bibnamefont{van Elp}},
  \bibinfo{author}{\bibfnamefont{P.}~\bibnamefont{Kuiper}},
  \bibinfo{author}{\bibfnamefont{G.~A.} \bibnamefont{Sawatzky}},
  \bibinfo{author}{\bibfnamefont{S.}~\bibnamefont{Hosoya}}, \bibnamefont{and}
  \bibinfo{author}{\bibfnamefont{H.}~\bibnamefont{Katayama-Yoshida}},
  \bibinfo{journal}{Phys. Rev. B} \textbf{\bibinfo{volume}{45}},
  \bibinfo{pages}{12513} (\bibinfo{year}{1992}).

\bibitem[{\citenamefont{Anisimov et~al.}(1992)\citenamefont{Anisimov, Korotin,
  Zaanen, and Andersen}}]{Anisimov92}
\bibinfo{author}{\bibfnamefont{V.~I.} \bibnamefont{Anisimov}},
  \bibinfo{author}{\bibfnamefont{M.~A.} \bibnamefont{Korotin}},
  \bibinfo{author}{\bibfnamefont{J.}~\bibnamefont{Zaanen}}, \bibnamefont{and}
  \bibinfo{author}{\bibfnamefont{O.~K.} \bibnamefont{Andersen}},
  \bibinfo{journal}{Phys. Rev. Lett.} \textbf{\bibinfo{volume}{68}},
  \bibinfo{pages}{345} (\bibinfo{year}{1992}).

\bibitem[{\citenamefont{Bi and Eklund}(1993)}]{Bi93}
\bibinfo{author}{\bibfnamefont{X.-X.} \bibnamefont{Bi}} \bibnamefont{and}
  \bibinfo{author}{\bibfnamefont{P.~C.} \bibnamefont{Eklund}},
  \bibinfo{journal}{Phys. Rev. Lett.} \textbf{\bibinfo{volume}{70}},
  \bibinfo{pages}{2625} (\bibinfo{year}{1993}).

\bibitem[{\citenamefont{Chen et~al.}(1993)\citenamefont{Chen, Cheong, and
  Cooper}}]{Chen93}
\bibinfo{author}{\bibfnamefont{C.~H.} \bibnamefont{Chen}},
  \bibinfo{author}{\bibfnamefont{S.-W.} \bibnamefont{Cheong}},
  \bibnamefont{and} \bibinfo{author}{\bibfnamefont{A.~S.}
  \bibnamefont{Cooper}}, \bibinfo{journal}{Phys. Rev. Lett.}
  \textbf{\bibinfo{volume}{71}}, \bibinfo{pages}{2461} (\bibinfo{year}{1993}).

\bibitem[{\citenamefont{Zaanen and Littlewood}(1994)}]{Zaanen94}
\bibinfo{author}{\bibfnamefont{J.}~\bibnamefont{Zaanen}} \bibnamefont{and}
  \bibinfo{author}{\bibfnamefont{P.~B.} \bibnamefont{Littlewood}},
  \bibinfo{journal}{Phys. Rev. B} \textbf{\bibinfo{volume}{50}},
  \bibinfo{pages}{7222} (\bibinfo{year}{1994}).

\bibitem[{\citenamefont{Cheong et~al.}(1994)\citenamefont{Cheong, Hwang, Chen,
  Batlogg, Rupp, and Carter}}]{Cheong94}
\bibinfo{author}{\bibfnamefont{S.-W.} \bibnamefont{Cheong}},
  \bibinfo{author}{\bibfnamefont{H.~Y.} \bibnamefont{Hwang}},
  \bibinfo{author}{\bibfnamefont{C.~H.} \bibnamefont{Chen}},
  \bibinfo{author}{\bibfnamefont{B.}~\bibnamefont{Batlogg}},
  \bibinfo{author}{\bibfnamefont{L.~W.} \bibnamefont{Rupp}}, \bibnamefont{and}
  \bibinfo{author}{\bibfnamefont{S.~A.} \bibnamefont{Carter}},
  \bibinfo{journal}{Phys. Rev. B} \textbf{\bibinfo{volume}{49}},
  \bibinfo{pages}{7088} (\bibinfo{year}{1994}).

\bibitem[{\citenamefont{Birgeneau et~al.}(1987)\citenamefont{Birgeneau, Chen,
  Gabbe, Jenssen, Kastner, Peters, Picone, Thio, Thurston, and
  Tuller}}]{Birgeneau87}
\bibinfo{author}{\bibfnamefont{R.~J.} \bibnamefont{Birgeneau}},
  \bibinfo{author}{\bibfnamefont{C.~Y.} \bibnamefont{Chen}},
  \bibinfo{author}{\bibfnamefont{D.~R.} \bibnamefont{Gabbe}},
  \bibinfo{author}{\bibfnamefont{H.~P.} \bibnamefont{Jenssen}},
  \bibinfo{author}{\bibfnamefont{M.~A.} \bibnamefont{Kastner}},
  \bibinfo{author}{\bibfnamefont{C.~J.} \bibnamefont{Peters}},
  \bibinfo{author}{\bibfnamefont{P.~J.} \bibnamefont{Picone}},
  \bibinfo{author}{\bibfnamefont{T.}~\bibnamefont{Thio}},
  \bibinfo{author}{\bibfnamefont{T.~R.} \bibnamefont{Thurston}},
\bibinfo{author}{\bibfnamefont{H.~L.} \bibnamefont{Tuller}},
\bibinfo{author}{\bibfnamefont{J.~D.} \bibnamefont{Axe}},
\bibinfo{author}{\bibfnamefont{P.} \bibnamefont{B\"{o}ni}},
  \bibnamefont{and} \bibinfo{author}{\bibfnamefont{G.}
  \bibnamefont{Shirane}}, \bibinfo{journal}{Phys. Rev. Lett.}
  \textbf{\bibinfo{volume}{59}}, \bibinfo{pages}{1329} (\bibinfo{year}{1987}).

\bibitem[{\citenamefont{Pintschovius}(1990)}]{Pintschovius90}
\bibinfo{author}{\bibfnamefont{L.}~\bibnamefont{Pintschovius}},
  \bibinfo{journal}{Festk\"orperprobleme} \textbf{\bibinfo{volume}{30}},
  \bibinfo{pages}{183} (\bibinfo{year}{1990}).

\bibitem[{\citenamefont{Mart{\'{\i}}nez
  et~al.}(1991)\citenamefont{Mart{\'{\i}}nez, Fern\'{a}ndez-D{\'{\i}}az,
  Rodr{\'{\i}}guez-Carvajal, and Odier}}]{Martinez91}
\bibinfo{author}{\bibfnamefont{J.~L.} \bibnamefont{Mart{\'{\i}}nez}},
  \bibinfo{author}{\bibfnamefont{M.~T.}
  \bibnamefont{Fern\'{a}ndez-D\'{\i}az}},
  \bibinfo{author}{\bibfnamefont{J.}~\bibnamefont{Rodr{\'{\i}}guez-Carvajal}},
  \bibnamefont{and} \bibinfo{author}{\bibfnamefont{P.}~\bibnamefont{Odier}},
  \bibinfo{journal}{Phys. Rev. B} \textbf{\bibinfo{volume}{43}},
  \bibinfo{pages}{13766} (\bibinfo{year}{1991}).

\bibitem[{\citenamefont{Keimer et~al.}(1993)\citenamefont{Keimer, Birgeneau,
  Cassanho, Endoh, Greven, Kastner, and Shirane}}]{Keimer93}
\bibinfo{author}{\bibfnamefont{B.}~\bibnamefont{Keimer}},
  \bibinfo{author}{\bibfnamefont{R.~J.} \bibnamefont{Birgeneau}},
  \bibinfo{author}{\bibfnamefont{A.}~\bibnamefont{Cassanho}},
  \bibinfo{author}{\bibfnamefont{Y.}~\bibnamefont{Endoh}},
  \bibinfo{author}{\bibfnamefont{M.}~\bibnamefont{Greven}},
  \bibinfo{author}{\bibfnamefont{M.~A.} \bibnamefont{Kastner}},
  \bibnamefont{and} \bibinfo{author}{\bibfnamefont{G.}~\bibnamefont{Shirane}},
  \bibinfo{journal}{Z. Phys. B} \textbf{\bibinfo{volume}{91}},
  \bibinfo{pages}{373} (\bibinfo{year}{1993}).

\bibitem[{\citenamefont{Sera et~al.}(1997)\citenamefont{Sera, Maki, Hiroi, and
  Kobayashi}}]{Sera97}
\bibinfo{author}{\bibfnamefont{M.}~\bibnamefont{Sera}},
  \bibinfo{author}{\bibfnamefont{M.}~\bibnamefont{Maki}},
  \bibinfo{author}{\bibfnamefont{M.}~\bibnamefont{Hiroi}}, \bibnamefont{and}
  \bibinfo{author}{\bibfnamefont{N.}~\bibnamefont{Kobayashi}},
  \bibinfo{journal}{J. Phys. Soc. Jpn.} \textbf{\bibinfo{volume}{66}},
  \bibinfo{pages}{765} (\bibinfo{year}{1997}).

%\bibitem[{\citenamefont{{Hess \em et al.}}(to be published)}]{Hess03tobe}
%\bibinfo{author}{\bibfnamefont{C.}~\bibnamefont{{Hess \em et al.}}}
%  (\bibinfo{year}{to be published}).



\bibitem{foot1}{IMI 7031 Insulating Varnish.}

\bibitem[{\citenamefont{Sch{\"a}fer et~al.}(1994)\citenamefont{Sch{\"a}fer,
  Breuer, Bauer, Freimuth, Knauf, Roden, Schlabitz, and
  B\"{u}chner}}]{Schafer94}
\bibinfo{author}{\bibfnamefont{W.}~\bibnamefont{Sch{\"a}fer}},
  \bibinfo{author}{\bibfnamefont{M.}~\bibnamefont{Breuer}},
  \bibinfo{author}{\bibfnamefont{G.}~\bibnamefont{Bauer}},
  \bibinfo{author}{\bibfnamefont{A.}~\bibnamefont{Freimuth}},
  \bibinfo{author}{\bibfnamefont{N.}~\bibnamefont{Knauf}},
  \bibinfo{author}{\bibfnamefont{B.}~\bibnamefont{Roden}},
  \bibinfo{author}{\bibfnamefont{W.}~\bibnamefont{Schlabitz}},
  \bibnamefont{and}
  \bibinfo{author}{\bibfnamefont{B.}~\bibnamefont{B\"{u}chner}},
  \bibinfo{journal}{Phys. Rev. B} \textbf{\bibinfo{volume}{49}},
  \bibinfo{pages}{9248} (\bibinfo{year}{1994}).

\bibitem[{\citenamefont{Nakamura and Uchida}(1992)}]{Nakamura92}
\bibinfo{author}{\bibfnamefont{Y.}~\bibnamefont{Nakamura}} \bibnamefont{and}
  \bibinfo{author}{\bibfnamefont{S.}~\bibnamefont{Uchida}},
  \bibinfo{journal}{Phys. Rev. B} \textbf{\bibinfo{volume}{46}},
  \bibinfo{pages}{5841} (\bibinfo{year}{1992}).

\bibitem[{\citenamefont{Nakamura and Uchida}(1993)}]{Nakamura93}
\bibinfo{author}{\bibfnamefont{Y.}~\bibnamefont{Nakamura}} \bibnamefont{and}
  \bibinfo{author}{\bibfnamefont{S.}~\bibnamefont{Uchida}},
  \bibinfo{journal}{Phys. Rev. B} \textbf{\bibinfo{volume}{47}},
  \bibinfo{pages}{8369} (\bibinfo{year}{1993}).

\bibitem[{\citenamefont{Thio et~al.}(1988)\citenamefont{Thio, Thurston, Preyer,
  Picone, Kastner, Jenssen, Gabbe, Chen, Birgeneau, and Aharony}}]{Thio88}
\bibinfo{author}{\bibfnamefont{T.}~\bibnamefont{Thio}},
  \bibinfo{author}{\bibfnamefont{T.~R.} \bibnamefont{Thurston}},
  \bibinfo{author}{\bibfnamefont{N.~W.} \bibnamefont{Preyer}},
  \bibinfo{author}{\bibfnamefont{P.~J.} \bibnamefont{Picone}},
  \bibinfo{author}{\bibfnamefont{M.~A.} \bibnamefont{Kastner}},
  \bibinfo{author}{\bibfnamefont{H.~P.} \bibnamefont{Jenssen}},
  \bibinfo{author}{\bibfnamefont{D.~R.} \bibnamefont{Gabbe}},
  \bibinfo{author}{\bibfnamefont{C.~Y.} \bibnamefont{Chen}},
  \bibinfo{author}{\bibfnamefont{R.~J.} \bibnamefont{Birgeneau}},
  \bibnamefont{and} \bibinfo{author}{\bibfnamefont{A.}~\bibnamefont{Aharony}},
  \bibinfo{journal}{Phys. Rev. B} \textbf{\bibinfo{volume}{38}},
  \bibinfo{pages}{905} (\bibinfo{year}{1988}).

\bibitem[{\citenamefont{Berman and Brock}(1965)}]{Berman65}
\bibinfo{author}{\bibfnamefont{R.}~\bibnamefont{Berman}} \bibnamefont{and}
  \bibinfo{author}{\bibfnamefont{J.}~\bibnamefont{Brock}},
  \bibinfo{journal}{Proc. Roy. Soc. (London)} \textbf{\bibinfo{volume}{A289}},
  \bibinfo{pages}{46} (\bibinfo{year}{1965}).

\bibitem[{\citenamefont{Pintschovius and Reichardt}(1998)}]{Pintschovius98}
\bibinfo{author}{\bibfnamefont{L.}~\bibnamefont{Pintschovius}}
  \bibnamefont{and}
  \bibinfo{author}{\bibfnamefont{W.}~\bibnamefont{Reichardt}}, in
  \emph{\bibinfo{booktitle}{Neutron Scattering in Layered Copper-Oxide
  Superconductors}}, edited by
  \bibinfo{editor}{\bibfnamefont{A.}~\bibnamefont{Furrer}}
  (\bibinfo{publisher}{Kluwer Academic Publishers},
  \bibinfo{address}{Dordrecht}, \bibinfo{year}{1998}), pp.
  \bibinfo{pages}{165--224}.

\bibitem[{\citenamefont{Steigmeier}(1968)}]{Steigmeier68}
\bibinfo{author}{\bibfnamefont{E.~F.} \bibnamefont{Steigmeier}},
  \bibinfo{journal}{Phys. Rev.} \textbf{\bibinfo{volume}{168}},
  \bibinfo{pages}{523} (\bibinfo{year}{1968}).

\bibitem[{\citenamefont{Barret and Holland}(1970)}]{Barret70}
\bibinfo{author}{\bibfnamefont{H.~H.} \bibnamefont{Barret}} \bibnamefont{and}
  \bibinfo{author}{\bibfnamefont{M.~G.} \bibnamefont{Holland}},
  \bibinfo{journal}{Phys. Rev. B} \textbf{\bibinfo{volume}{2}},
  \bibinfo{pages}{3441} (\bibinfo{year}{1970}).

\bibitem{foot2}{We should note that Sera et al.
explain the change of $\kappa_{\mathrm{ph}}$ at $T_{LT}$ by an observed change of the
velocity of sound $v_s$ at $T_{LT}$.\cite{Sera97} The actual changes of $v_s$ are, however, far too
small ($\sim1\%$)\cite{Yamada94} to account for the much larger changes of $\kappa_{\mathrm{ph}}$.
We therefore regard these changes of $v_s$ at $T_{LT}$ as a further accompanying phenomenon of the
structural phase transition.}

\bibitem[{\citenamefont{Yamada et~al.}(1994)\citenamefont{Yamada, Sera, Sato,
  Takayama, Takata, and Sakata}}]{Yamada94}
\bibinfo{author}{\bibfnamefont{J.}~\bibnamefont{Yamada}},
  \bibinfo{author}{\bibfnamefont{M.}~\bibnamefont{Sera}},
  \bibinfo{author}{\bibfnamefont{M.}~\bibnamefont{Sato}},
  \bibinfo{author}{\bibfnamefont{T.}~\bibnamefont{Takayama}},
  \bibinfo{author}{\bibfnamefont{M.}~\bibnamefont{Takata}}, \bibnamefont{and}
  \bibinfo{author}{\bibfnamefont{M.}~\bibnamefont{Sakata}},
  \bibinfo{journal}{J. Phys. Soc. Jpn.} \textbf{\bibinfo{volume}{63}},
  \bibinfo{pages}{2314} (\bibinfo{year}{1994}).

\bibitem{foot2a}Note that the reported degeneracy of energy of acoustic and soft optical phonons is $\sim 10$~meV.\cite{Birgeneau87} Hence, the effect of soft-phonon scattering on $\kappa_{\mathrm{ph}}$ should be strongest around $\sim 25$~K, i.e., in the vicinity of the phononic peak, since acoustic phonons with energy $\sim 4k_BT$ contribute most significantly to $\kappa_{\mathrm{ph}}$.\cite{Berman} Apart from this lowest-order scattering, three-phonon processes and processes of higher order involving much softer phonon modes close to the Brillouin zone center with energies between 2meV and 5meV\cite{Birgeneau87,Keimer93,Braden94} are very likely.


\bibitem[{\citenamefont{Braden et~al.}(1994)\citenamefont{Braden, Schnelle, Schwarz, Pyka, Heger, Fisk, Gamayunov, Tanaka, and Kojima }}]{Braden94}
\bibinfo{author}{\bibfnamefont{M.}~\bibnamefont{Braden}},
  \bibinfo{author}{\bibfnamefont{W.} \bibnamefont{Schnelle}},
  \bibinfo{author}{\bibfnamefont{W.}~\bibnamefont{Schwarz}},
  \bibinfo{author}{\bibfnamefont{N.}~\bibnamefont{Pyka}},
  \bibinfo{author}{\bibfnamefont{G.}~\bibnamefont{Heger}},
  \bibinfo{author}{\bibfnamefont{Z.} \bibnamefont{Fisk}},
  \bibinfo{author}{\bibfnamefont{K.} \bibnamefont{Gamayunov}},
  \bibinfo{author}{\bibfnamefont{I.} \bibnamefont{Tanaka}},
  \bibnamefont{and} \bibinfo{author}{\bibfnamefont{H.}~\bibnamefont{Kojima}},
  \bibinfo{journal}{Z. Phys. B} \textbf{\bibinfo{volume}{94}},
  \bibinfo{pages}{29} (\bibinfo{year}{1994}).


\bibitem[{\citenamefont{Berman}(1976)}]{Berman}
\bibinfo{author}{\bibfnamefont{R.}~\bibnamefont{Berman}},
  \emph{\bibinfo{title}{Thermal Conduction in Solids}} (\bibinfo{publisher}{At
  the Clarendon Press, Oxford}, \bibinfo{year}{1976}).


\bibitem{foot3}{Note that the elastic constants of this material are
only weakly anisotropic.\cite{Pintschovius91}}


\bibitem[{\citenamefont{Pintschovius et~al.}(1991)\citenamefont{Pintschovius,
  Pyka, Reichardt, Rumiantsev, Mitrofanov, Ivanov, Collin, and
  Bourges}}]{Pintschovius91}
\bibinfo{author}{\bibfnamefont{L.}~\bibnamefont{Pintschovius}},
  \bibinfo{author}{\bibfnamefont{N.}~\bibnamefont{Pyka}},
  \bibinfo{author}{\bibfnamefont{W.}~\bibnamefont{Reichardt}},
  \bibinfo{author}{\bibfnamefont{A.~Y.} \bibnamefont{Rumiantsev}},
  \bibinfo{author}{\bibfnamefont{N.~L.} \bibnamefont{Mitrofanov}},
  \bibinfo{author}{\bibfnamefont{A.~S.} \bibnamefont{Ivanov}},
  \bibinfo{author}{\bibfnamefont{G.}~\bibnamefont{Collin}}, \bibnamefont{and}
  \bibinfo{author}{\bibfnamefont{P.}~\bibnamefont{Bourges}},
  \bibinfo{journal}{Physica C} \textbf{\bibinfo{volume}{185-189}},
  \bibinfo{pages}{156} (\bibinfo{year}{1991}).

\bibitem{foot4}{The low-$T$ energy of the soft mode at the zone center decreases from 2.44~meV ($x=0$) to 1.82~meV ($x=0.13$).\cite{Braden94}}


\bibitem[{\citenamefont{Pintschovius et~al.}(1989)\citenamefont{Pintschovius,
  Bassat, Odier, Gervais, Chevrier, Reichardt, and Gompf}}]{Pintschovius89}
\bibinfo{author}{\bibfnamefont{L.}~\bibnamefont{Pintschovius}},
  \bibinfo{author}{\bibfnamefont{J.~M.} \bibnamefont{Bassat}},
  \bibinfo{author}{\bibfnamefont{P.}~\bibnamefont{Odier}},
  \bibinfo{author}{\bibfnamefont{F.}~\bibnamefont{Gervais}},
  \bibinfo{author}{\bibfnamefont{G.}~\bibnamefont{Chevrier}},
  \bibinfo{author}{\bibfnamefont{W.}~\bibnamefont{Reichardt}},
  \bibnamefont{and} \bibinfo{author}{\bibfnamefont{F.}~\bibnamefont{Gompf}},
  \bibinfo{journal}{Phys. Rev. B} \textbf{\bibinfo{volume}{40}},
  \bibinfo{pages}{2229} (\bibinfo{year}{1989}).

\bibitem[{\citenamefont{Pintschovius et~al.}(2001)\citenamefont{Pintschovius,
  Reichardt, Braden, Dhalenne, and Revcolevschi}}]{Pintschovius01}
\bibinfo{author}{\bibfnamefont{L.}~\bibnamefont{Pintschovius}},
  \bibinfo{author}{\bibfnamefont{W.}~\bibnamefont{Reichardt}},
  \bibinfo{author}{\bibfnamefont{M.}~\bibnamefont{Braden}},
  \bibinfo{author}{\bibfnamefont{G.}~\bibnamefont{Dhalenne}}, \bibnamefont{and}
  \bibinfo{author}{\bibfnamefont{A.}~\bibnamefont{Revcolevschi}},
  \bibinfo{journal}{Phys. Rev. B} \textbf{\bibinfo{volume}{64}},
  \bibinfo{pages}{094510} (\bibinfo{year}{2001}).

\bibitem[{\citenamefont{Rodr{\'{\i}}guez-Carvajal
  et~al.}(1988)\citenamefont{Rodr{\'{\i}}guez-Carvajal, Mart{\'{\i}}nez,
  Pannetier, and Saez-Puche}}]{Rodriguez88}
\bibinfo{author}{\bibfnamefont{J.}~\bibnamefont{Rodr{\'{\i}}guez-Carvajal}},
  \bibinfo{author}{\bibfnamefont{J.~L.} \bibnamefont{Mart{\'{\i}}nez}},
  \bibinfo{author}{\bibfnamefont{J.}~\bibnamefont{Pannetier}},
  \bibnamefont{and}
  \bibinfo{author}{\bibfnamefont{R.}~\bibnamefont{Saez-Puche}},
  \bibinfo{journal}{Phys. Rev. B} \textbf{\bibinfo{volume}{38}},
  \bibinfo{pages}{7148} (\bibinfo{year}{1988}).


\bibitem{foot5}We note that similar as in $\rm
La_2CuO_4$, magnetic contributions might be present for $T\gtrsim100\rm ~K$.

\bibitem[{\citenamefont{Takagi et~al.}(1992)\citenamefont{Takagi, Cava,
  Marezio, Batlogg, Krajewski, W.~F.~Peck, Bordet, and Cox}}]{Takagi92a}
\bibinfo{author}{\bibfnamefont{H.}~\bibnamefont{Takagi}},
  \bibinfo{author}{\bibfnamefont{R.~J.} \bibnamefont{Cava}},
  \bibinfo{author}{\bibfnamefont{M.}~\bibnamefont{Marezio}},
  \bibinfo{author}{\bibfnamefont{B.}~\bibnamefont{Batlogg}},
  \bibinfo{author}{\bibfnamefont{J.~J.} \bibnamefont{Krajewski}},
  \bibinfo{author}{\bibfnamefont{W.~F.}~\bibnamefont{Peck}},
  \bibinfo{author}{\bibfnamefont{P.}~\bibnamefont{Bordet}}, \bibnamefont{and}
  \bibinfo{author}{\bibfnamefont{D.~E.} \bibnamefont{Cox}},
  \bibinfo{journal}{Phys. Rev. Lett.} \textbf{\bibinfo{volume}{68}},
  \bibinfo{pages}{3777} (\bibinfo{year}{1992}).

\bibitem[{\citenamefont{Hess and B\"{u}chner}(2003)}]{Hess03b}
\bibinfo{author}{\bibfnamefont{C.}~\bibnamefont{Hess}} \bibnamefont{and}
  \bibinfo{author}{\bibfnamefont{B.}~\bibnamefont{B\"{u}chner}},
  \bibinfo{journal}{cond-mat/0304248 (unpublished)}.

\end{thebibliography}

\bibliographystyle{apsrev}
\end{document}